\documentclass[a4paper,fleqn,usenatbib]{mnras}

\usepackage{newtxtext,newtxmath}

\usepackage[T1]{fontenc}
\usepackage{ae,aecompl}

\usepackage{graphicx}	
\usepackage{amsmath}	
\usepackage{amssymb}	

\usepackage{color}

\definecolor{gr}{cmyk}{1,0,1,0.09}

\title[SRP Resonances in LEO]{Solar Radiation Pressure Resonances in Low Earth Orbits}

\author[E. M. Alessi et al.]{
Elisa Maria Alessi,$^{1}$\thanks{E-mail: em.alessi@ifac.cnr.it (EMA)}
G. Schettino,$^{1}$
A. Rossi$^{1}$
and G. B. Valsecchi$^{1,2}$
\\
$^{1}$IFAC-CNR, via Madonna del Piano 10, 50019 Sesto Fiorentino (FI), Italy\\
$^{2}$IAPS-INAF, via Fosso del Cavaliere 100, 00133 Rome, Italy
}

\date{Accepted XXX. Received YYY; in original form ZZZ}

\pubyear{2017}

\begin{document}
\label{firstpage}
\pagerange{\pageref{firstpage}--\pageref{lastpage}}
\maketitle

\begin{abstract}

  The aim of this work is to highlight the crucial role that orbital
  resonances associated with solar radiation pressure can have in Low
  Earth Orbit. We review the corresponding literature, and provide an
  analytical tool to estimate the maximum eccentricity which
  can be achieved for well-defined initial conditions. We then compare
  the results obtained with the simplified model with the results
  obtained with a more
  comprehensive dynamical model. The analysis has important
  implications both from a theoretical point of view, because it shows
  that the role of some resonances was underestimated in the past, but
  also from a practical point of view in the perspective of passive
  deorbiting solutions for satellites at the end-of-life.

\end{abstract}

\begin{keywords}
Solar Radiation Pressure -- Resonances -- Low Earth Orbits
\end{keywords}



\section{Introduction}

It is known that the effect of the solar radiation pressure (SRP) on
the orbital motion is a long-period variation in eccentricity $e$ and
inclination $i$, along with one in longitude of ascending node
$\Omega$ and argument of pericenter $\omega$, and that the magnitude
of this variation depends on the area-to-mass ratio of the body
\citep[e.g.,][]{B2005}. If we assume a perturbing potential which is
averaged over the orbital period of the body, the SRP effect can be
subject to resonances, associated with a commensurability among the
rate of precession of the ascending node, the one of the argument of
pericenter and the mean apparent motion of the Sun, say $n_S$. As
well-known, when a resonance occurs, the period of the variation can
become so large to give rise to quasi-secular effects in the
corresponding orbital element.

Motivated by the numerical results \citep{A2017a,A2017b} obtained
recently within the ReDSHIFT H2020 project \citep{aR17}, the aim of
this work is to describe the role of SRP resonances in the Low Earth Orbit (LEO) region, showing that
their contribution can provide relatively significant eccentricity
variations.

In the past, the effect of the solar radiation pressure for orbits
around the Earth was considered mainly in the perspective of bodies
characterised by a very high area-to-mass ratio. We can mention works
focused on the behaviour of Geostationary Earth Orbits (GEO)
\citep[e.g.,][]{VetAl2007, RS2013, CetAl2015}, or on mission concepts
aiming to exploit a solar sail to deorbiting non-operational objects from Medium Earth Orbits (MEO)
\citep[e.g.,][]{LetAl2012}, or on the design of frozen orbits for
small-size objects \citep[e.g.,][]{CetAl2012, L15}. \cite{CetAl2012},
in particular, analysed SRP effects on high area-to-mass objects in order to
design frozen orbits for a swarm of small spacecraft dedicated to Earth
observation and telecommunication purposes. Starting from the work by
\cite{K96}, they studied the behaviour in eccentricity corresponding to
$\dot\Omega+\dot\omega-n_S=0$, by computing the equilibrium points, and
their stability, in a dynamical system including the oblateness of the
Earth and SRP.

From a theoretical point of view, the first works on the resonances
induced by SRP started from the analysis of the resonances induced by
the solar gravitational attraction. As a matter of fact, the two
disturbing potentials differ in the first order of the expansion and
in the amplitude of the perturbation \citep{H77}. If we focus on the resonances affecting the eccentricity evolution, \cite{pM60} and
\cite{gC62} were the first to highlight the existence of six SRP
resonances, and to show their location in the $(i,a)$
plane. \cite{gC62}, in particular, noted that, contrary to lunisolar
gravitational resonances, the SRP resonances are able to give a
variation in eccentricity also in the case of circular
orbits. \cite{pM60} labelled $\dot\Omega+\dot\omega-n_S$ as the `most
interesting resonance', as done by all the following
authors. In the case of the gravitational perturbation, this resonance is also known as {\it evection resonance}\footnote{To be precise, $2(\dot\Omega+\dot\omega-n_S)$.} \citep[e.g.,][]{BC61, TW1998, FetAl10}. \cite{sB99}, basing his analysis on the work done by
\cite{gC62}, noted that `resonant solar perturbations can be much
stronger than the lunar ones'. He also considered
$\dot\Omega+\dot\omega-n_S$ as the dominant resonance for prograde
orbits and $\dot\Omega-\dot\omega+n_S$ for retrograde orbits. In his
work, the strength of a given resonance is determined by the magnitude
of the libration region. As already mentioned, a fundamental
contribution to the topic was given by Hughes \citep{H77, H80, H81},
who tried to provide a more general treatment to the problem, by
analysing also the effect due to high-order terms\footnote{both in the
  eccentricity of the body and in the eccentricity of the mean
  apparent solar motion.}. He stressed that SRP resonances give rise
to variations not only in eccentricity, but also in inclination,
longitude of the ascending node and argument of pericenter. He
considered a resonance as dominant according to the magnitude of the
corresponding amplitude, function of semi-major axis $a$ and
eccentricity. More recently, \cite{C17} recognised the occurrence of
semi-secular resonances, as the SRP ones, in the LEO
region, but they stated that `Since the semi-secular resonances occur
at specific altitudes, the width of such resonances is small, and
since the air drag provokes a decay of the orbits on relatively short
time scales, one expects that these resonances play a minor role in
the long-term evolution of space debris'. We will show that this is
partially true. Though the regions where SRP resonances are effective are indeed narrow, for high values of area-to-mass ratio, they represent  the main mechanism to drive a satellite back to the Earth also when the LEO is sufficiently high to have a 
negligible interaction with the residual atmosphere, e.g., above an
altitude $h\simeq 1000$ km. On the other hand, if we consider low values of area-to-mass ratio, the combined effect of the atmospheric drag and the SRP resonances can reduce significantly the lifetime of the object.

One of the fundamental issue to mitigate the space debris problem and
control the growth of the debris population is to see if there exist natural
perturbations which facilitate the reentry to the Earth or, when this is not
possible, ensure a stable graveyard orbit in the long-term. At
altitudes when the atmospheric drag is not effective, the only way to
lower the pericenter altitude is to obtain an eccentricity
increase. For MEO, like those of the GNSS
constellations, the gravitational lunisolar perturbations can help to
this end too \citep{aR15, A16}. In LEO, apart from the atmospheric drag, the main natural effects that can be exploited are the lunisolar gravitational resonance, in particular the one corresponding to $\dot\Omega+2\dot\omega$ \citep{A2017a,A2017b} and the SRP. In particular, as just mentioned, area-to-mass values representing small
spacecraft equipped with a solar sail can experience a change as
dramatic as to reenter to the Earth, even without the support of the
atmospheric drag. These situations occur in correspondence of the
$\dot\Omega+\dot\omega-n_S$ resonance, considered as dominant in the
past, but not only. On the other hand, 
understanding how the eccentricity might change due to the SRP is crucial also to
control the stability of a graveyard orbit.

As a final note, we acknowledge that looking for the values of
area-to-mass ratio ensuring a reentry, given initial semi-major axis
and inclination, \cite{cC16} found numerically the same reentry
corridors represented by at least two of the six SRP resonances, that will be
described here.

\section{The Model}

Let us assume that the radiation coming from the Sun is directed
normally to the surface of the spacecraft, that is, the so-called
cannonball model. Moreover, let us assume the orbit of the spacecraft
entirely in the sunlight, and the effect of Earth's albedo as
negligible.

Neglecting the light aberration, the solar radiation pressure is a
conservative force. The corresponding disturbing potential, averaged
over the orbital motion of the spacecraft, can be written as
\citep[e.g.,][]{K96}
\begin{eqnarray}\nonumber
{\mathcal{R}}_{SRP}&=&\frac{3}{2}PC_R\frac{A}{m}ae[
\cos\omega\left(\cos\Omega\cos\lambda_S+\sin\Omega\sin\lambda_S\cos\epsilon\right)\\\nonumber
&&+\sin\omega\left(\cos\Omega\cos i\sin\lambda_S\cos\epsilon-\sin\Omega\cos i\cos\lambda_S\right)
\\\nonumber
&&+\sin i\sin\omega\sin\lambda_S\sin\epsilon
],\\\label{SRP_disturbing_function}
&=&\frac{3}{2}PC_R\frac{A}{m}ae[\mathcal{T}_1\cos \psi_1+\mathcal{T}_2\cos \psi_2+\mathcal{T}_3\cos \psi_3+\\\nonumber
&&\mathcal{T}_4\cos \psi_4+\mathcal{T}_5\cos \psi_5+\mathcal{T}_6\cos \psi_6],
\end{eqnarray}
where $\lambda_S$ is the longitude of the Sun measured on the ecliptic plane, $\epsilon=23.439^{\circ}$ is the obliquity of the ecliptic, $P$ the solar radiation pressure , $C_R$ the reflectivity coefficient, $A/m$ the area-to-mass ratio,
\begin{equation}\label{psi}
\psi_j=n_1\Omega+n_2\omega + n_3 \lambda_S,
\end{equation}
with $n_1=\{0,1\}$, $n_2=\pm 1$, $n_3=\pm 1$, according to $j$, following Table~\ref{tab:dotpsij}, and 
\begin{eqnarray}\nonumber
\mathcal{T}_1&=&\cos ^2\left(\frac{\epsilon}{2}\right) \cos ^2\left(\frac{i}{2}\right) \\\nonumber
\mathcal{T}_2&=&\cos^2 \left(\frac{\epsilon}{2}\right) \sin ^2\left(\frac{i}{2}\right)\\\label{SRP_disturbing_function_resonant_terms}
\mathcal{T}_3&=&\frac{1}{2} \sin (\epsilon) \sin (i) \\\nonumber
\mathcal{T}_4&=&-\frac{1}{2} \sin (\epsilon) \sin (i)\\\nonumber
\mathcal{T}_5&=&\sin ^2\left(\frac{\epsilon}{2}\right) \cos ^2\left(\frac{i}{2}\right)\\\nonumber
\mathcal{T}_6&=&\sin ^2\left(\frac{\epsilon}{2}\right) \sin ^2\left(\frac{i}{2}\right).
\end{eqnarray}

\begin{table}
	\centering
	\caption{Resonance $\psi_j=n_1\Omega+n_2\omega + n_3 \lambda_S$ in terms of $n_1, n_2, n_3$. }
	\label{tab:dotpsij}
	\begin{tabular}{lccc} 
		\hline
		$j$ & $n_1$ & $n_2$ & $n_3$\\
		\hline
		1 & 1 & 1 & -1\\
		2 & 1 & -1 & -1\\
		3 & 0 & 1 & -1\\
		4 & 0 & 1 & 1\\
		5 & 1 & 1 & 1\\
		6 & 1 & -1 & 1\\
		\hline
	\end{tabular}
\end{table}

\begin{figure}
	\includegraphics[width=\columnwidth]{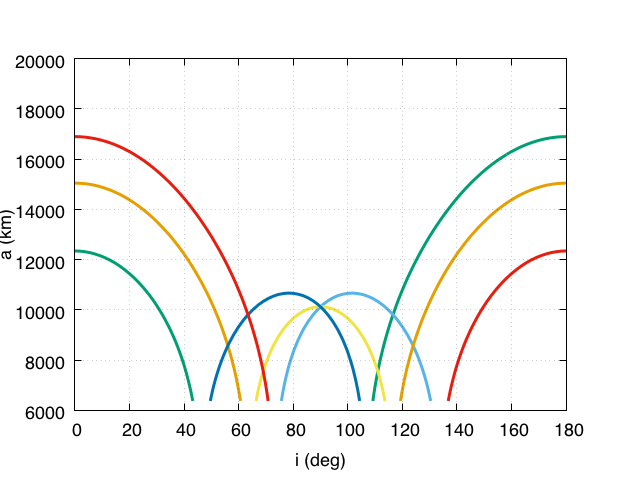}
	\includegraphics[width=\columnwidth]{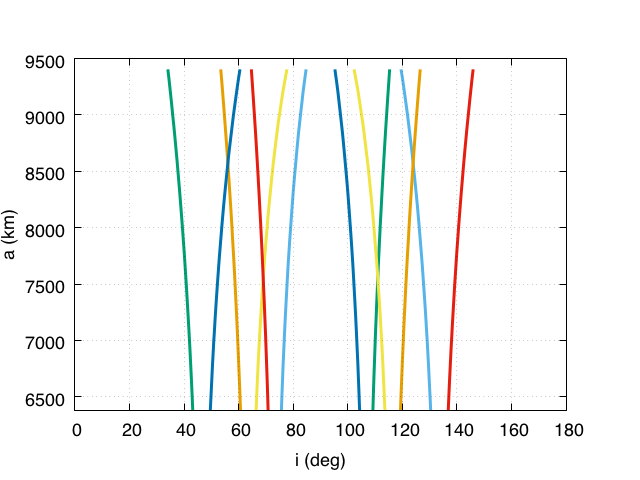}
    \caption{On the top, we show the location of the six main SRP resonances as a function of $i,a$ for $e=0.01$. On the bottom, a close-up in the LEO region, defined here up to $h=3000$ km, in order to account also for possible graveyard orbits. The curves were computed assuming $\dot\Omega=\dot\Omega_{J_2}$ and $\dot\omega=\dot\omega_{J_2}$. Green: $\dot\Omega+\dot\omega- n_{S}=0$. Cyan: $\dot\Omega-\dot\omega- n_{S}=0$. Orange: $\dot\omega- n_{S}=0$. Yellow: $\dot\omega+ n_{S}=0$. Blue: $\dot\Omega+\dot\omega+ n_{S}=0$. Red: $\dot\Omega-\dot\omega+ n_{S}=0$. }
   \label{fig:res}
\end{figure}

\begin{figure}
	\includegraphics[width=\columnwidth]{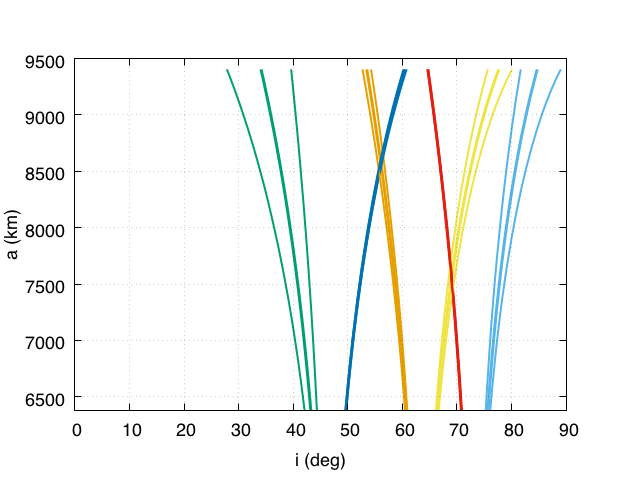}
	\includegraphics[width=\columnwidth]{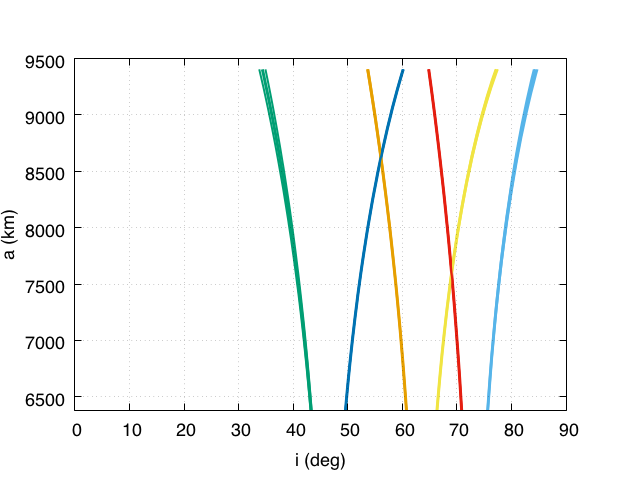}
    \caption{Location of the six main SRP resonances as a function of $i,a$ for $e=0.01$ (top) and $e=0.1$ (bottom) in the LEO region for prograde orbits for $A/m=1$ m$^2/$kg. The curves were computed assuming $\dot\Omega=\dot\Omega_{J_2}+\dot\Omega_{SRP}$ and $\dot\omega=\dot\omega_{J_2}+\dot\omega_{SRP}$. Green: $\dot\Omega+\dot\omega- n_{S}=0$.  Cyan: $\dot\Omega-\dot\omega- n_{S}=0$. Orange: $\dot\omega- n_{S}=0$. Yellow: $\dot\omega+ n_{S}=0$. Blue: $\dot\Omega+\dot\omega+ n_{S}=0$. Red: $\dot\Omega-\dot\omega+ n_{S}=0$. For a given resonance, the middle curve corresponds to $\psi=90^{\circ}$. If $n_3=1$ the right curve corresponds to $\psi=0^{\circ}$, the left one to $\psi=180^{\circ}$; if $n_3=-1$ viceversa.}
   \label{fig:res_con_SRP}
\end{figure}

\begin{figure*}
\includegraphics[width=0.66\columnwidth]{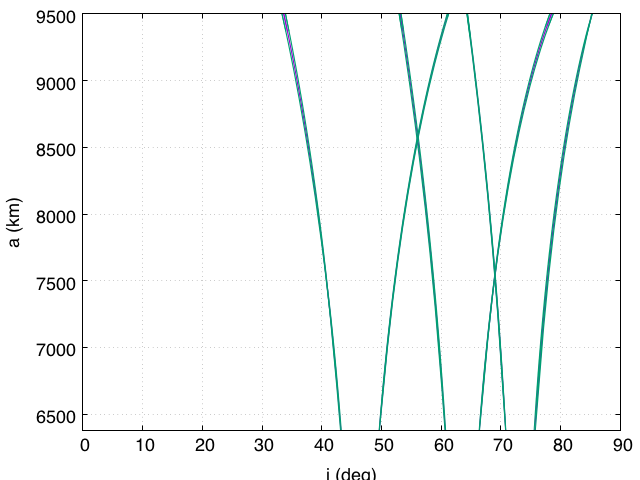}
\includegraphics[width=0.66\columnwidth]{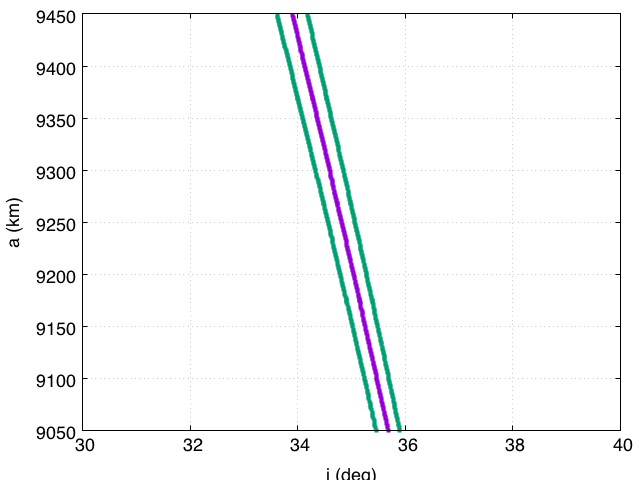}
\includegraphics[width=0.66\columnwidth]{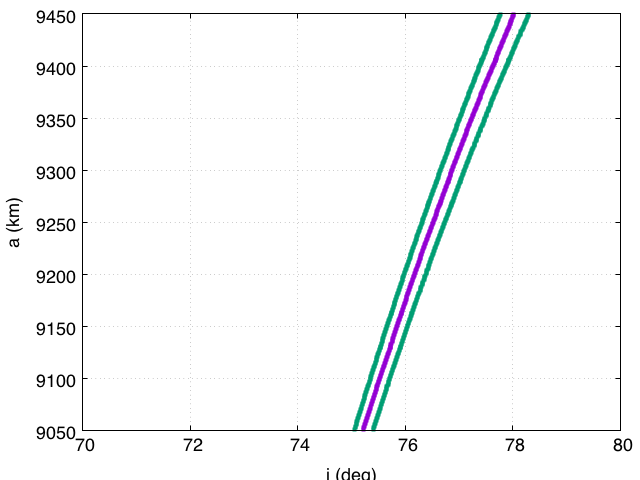}
    \caption{Resonance width, computed for $e=0.01$ and $A/m=1$ m$^2/$kg. The middle and right panels show the regions where the width computed is the largest: they correspond to resonance \#1 and resonance \#3, respectively.}
   \label{fig:width_high}
\end{figure*}

If we assume that the dynamics is
  driven by a single resonance, then we can write the variation in the orbital
elements as due to only one, say $j$, of the terms in
(\ref{SRP_disturbing_function}) at a time. By applying
the Lagrange planetary equations, we get
\begin{eqnarray}\label{equations}
\frac{de}{dt}&=&-\frac{3}{2}PC_R\frac{A}{m}\frac{\sqrt{1-e^2}}{na}\mathcal{T}_{j}{\partial{\cos{\psi}_j}\over\partial\omega}\\\nonumber
\frac{di}{dt}&=&-\frac{3}{2}PC_R\frac{A}{m}\frac{e}{na\sqrt{1-e^2}\sin i}\mathcal{T}_{j}\left({\partial{\cos{\psi}_j}\over\partial \Omega}-\cos i{\partial{\cos{\psi}_j}\over\partial \omega}\right)\\\nonumber
\frac{d\Omega}{dt}&=&\dot\Omega_{J_2}+\dot\Omega_{SRP}\\\nonumber
\frac{d\omega}{dt}&=&\dot\omega_{J_2}+\dot\omega_{SRP},\\\nonumber
\end{eqnarray}
where $n$ is the mean motion of the satellite and we have assumed that
the only other effect exerting a perturbation on the spacecraft is the
oblateness of the Earth. We have
\begin{eqnarray}\label{bompom}
\dot\Omega_{J_2} &=& -\frac{3}{2}\frac{J_2r^2_{\bigoplus}n}{a^2(1-e^2)^2}\cos i\\\nonumber
\dot\Omega_{SRP}&=&\frac{3}{2}PC_R\frac{A}{m}\frac{e}{na\sqrt{1-e^2}\sin i}{\partial{\mathcal{T}}_{j}\over\partial i}\cos{\psi}_j\\\nonumber
\dot\omega_{J_2} &=& \frac{3}{4}\frac{J_2r^2_{\bigoplus}n}{a^2(1-e^2)^2}\left(5\cos^2i-1\right)\\\nonumber
\dot\omega_{SRP}&=&\frac{3}{2}PC_R\frac{A}{m}\left(\frac{\sqrt{1-e^2}}{nae}\mathcal{T}_{j}\cos{\psi}_j-\frac{e\cos i}{na\sqrt{1-e^2}\sin i}{\partial{\mathcal{T}}_{j}\over\partial i}\cos{\psi}_j\right),
\end{eqnarray}
  where $J_2$ is the second zonal term
of the geopotential and $r_{\bigoplus}$ the equatorial radius of the
Earth. The effect of lunisolar perturbations on
$\dot\Omega, \dot\omega$ can be neglected, following, e.g.,
\cite{MNF87}.

Notice that long-period and secular effects associated with the oblateness of the Earth affect only the behaviour of the longitude of the ascending node, of the argument of pericenter and of the mean anomaly. That is, the eccentricity does not change in the long-term due to $J_2$ (see, e.g.,  \cite{Roy}).

\section{The Eccentricity Resonances}

From Eq.~(\ref{equations}), we can recognise the following 6 resonances affecting the behaviour in eccentricity
\begin{equation}\label{dotpsi}
\dot\psi_j=n_1\dot\Omega+n_2\dot\omega + n_3 n_{S}=0.
\end{equation}
In Fig.~\ref{fig:res} we show their location as a function of the
inclination and semi-major axis, for given values of $e$ and $A/m$, assuming that
only the oblateness of the Earth is responsible of a variation in
  $\Omega$ and $\omega$. This approximation can be considered
valid for $A/m=0.012$ m$^2/$kg, which represents the average value of
the intact objects orbiting in LEO. On the bottom panel, we show a close-up in the LEO region, defined here up to $h=3000$ km, in order to account also for possible graveyard orbits. Note that we prefer to represent the resonances as a function of $i,a$ instead of $a,e$, as done, e.g.,
in \cite{CM12}, because we focus on values of eccentricity which
represent orbits of operational spacecraft in LEO. Realistic values cannot
exceed the value $e=0.15$, the majority of the population in LEO
having actually $e<0.02$. In general, except in the cases where the spacecraft is orbiting in a region where the atmospheric drag is dominant, or at a given resonance affecting the eccentricity evolution, the eccentricity can be assumed to vary in a negligible way with respect to the initial value. Also, in case of resonance, the timescale variation of the eccentricity is much slower than the typical period of the orbits considered.

In Fig. ~\ref{fig:res_con_SRP}, we show the location of the same
resonances for two different values of eccentricity, accounting also
  for the variation of $\Omega$ and $\omega$ due to SRP. In this
case, the location of the resonances depends also on
$\Omega, \omega, \lambda_S$. In the Figure we compute the curves
corresponding to $\psi=0^{\circ}, 90^{\circ}, 180^{\circ}$ for $A/m=1$
m$^2/$kg which represents a feasible value, given the current level of
technology, for a small spacecraft equipped with a SRP enhancing
device (see \cite{Colombo_IAC17}). For a given resonance, the middle curve corresponds to the
case $\psi=90^{\circ}$, which is equivalent to the case of
Fig.~\ref{fig:res}. The displacement among the curves of a same
resonance can be appreciated at higher LEO, for quasi-circular
orbits for the first and the second resonance, following the
convention in Table~\ref{tab:dotpsij}. Notice, however, that the first equation in (\ref{equations}) states that $\psi=0^{\circ}$ and $\psi=180^{\circ}$ are stationary points for the eccentricity, and thus we expect to have a quasi-secular behaviour in eccentricity only at approaching the condition $\psi=90^{\circ}$. 

In the same Figures, we note that resonances of different nature can
cross each other. Focusing on prograde LEO, there exist two main
overlapping between SRP resonances. For quasi circular orbits, i.e.,
$e=0.01$, we have
\begin{itemize}
\item an overlapping between $\dot\psi=\dot\omega-n_S=0$ and $\dot\psi=\dot\Omega+\dot\omega+n_S=0$ at $a\approx r_{\bigoplus} + 2193$ km and $i\approx 56.06^{\circ}$\footnote{Note that at the same inclination it appears also the well-known lunisolar gravitational resonance $\dot\Omega+2\dot\omega$.};
\item an overlapping between $\dot\psi=\dot\omega+n_S=0$ and $\dot\psi=\dot\Omega-\dot\omega+n_S=0$ at $a\approx r_{\bigoplus} + 1180$ km and $i\approx 69^{\circ}$.
\end{itemize}
Increasing the eccentricity, the value of the semi-major axis at which
the crossing appears also increases, e.g., for $e=0.1$ the first
crossing occurs at $a\approx r_{\bigoplus} + 2245$ km, the second at
$a\approx r_{\bigoplus} + 1220$ km. From preliminary simulations, the
dynamics at the overlapping does not manifest a chaotic
behaviour, rather the two resonances concur to a possible increase in
eccentricity. In general, we can observe isolated points where the Lyapunov time is significantly lower than that in the neighbourhood. They do not affect the overall dynamics, but they can be, however, object of further dedicated studies.

\begin{figure*}
\includegraphics[width=\columnwidth]{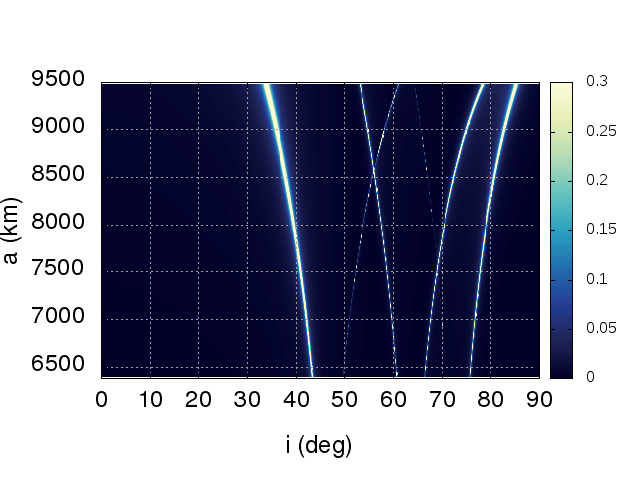}
\includegraphics[width=\columnwidth]{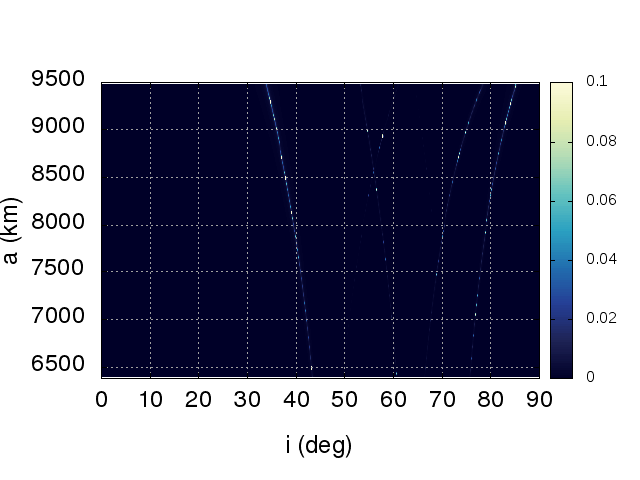}
    \caption{Maximum variation in eccentricity (color bar)  that can be obtained along a given SRP resonance as a function of $i,a$ for $e=0.01$, $\Omega=0^{\circ}$, $\omega=0^{\circ}$, $\lambda_S=90.086^{\circ}$. Left: $A/m=1$ m$^2/$kg. Right:  $A/m=0.012$ m$^2/$kg. See further details in the text.}
   \label{fig:delta_e}
\end{figure*}

To compute the width of the resonances considered, it can be applied the same argument developed by \cite{D16} for lunisolar doubly-averaged gravitational perturbations. Let us assume that the Hamiltonian function of the system is
\begin{equation}\label{eq:ham}
\mathcal{H}=\mathcal{H}_{K}+\mathcal{H}_{J_2}+\mathcal{H}_{SRP},
\end{equation}
where $\mathcal{H}_{K}$ is the Keplerian part, $\mathcal{H}_{J_2}$ the one associated with the oblateness of the Earth, and $\mathcal{H}_{SRP}=-{\mathcal{R}}_{SRP}$. The curves defining the boundaries of a given resonance $j$ at $X^*\equiv (e^*,i^*)$ are defined by the condition 
\begin{equation}\label{eq:width}
n_1\dot \Omega+n_2\dot \omega + n_3 n_S = \pm 2\sqrt{\nu}\left({\partial^2 \mathcal{H}^{sec}\over\partial X^2}\right)_{|X=X^*},
\end{equation}
where 
\begin{equation}\label{eq:nu_SRP}
\nu=\left|\frac{\frac{3}{2}PC_R\frac{A}{m}ae\mathcal{T}_j}{\left({\partial^2 \mathcal{H}^{sec}\over\partial X^2}\right)_{|X=X^*}}\right|,
\end{equation}
and ${\partial^2 \mathcal{H}^{sec}\over\partial X^2}$ as in \cite{D16}. For the sake of clarity\footnote{Note that $n_1$ are $n_2$ are exchanged in our formulation with respect to the formulation in \cite{D16}.}, 
$$
{\partial^2 \mathcal{H}^{sec}\over\partial X^2}=\frac{3}{2}\frac{J_2r^2_{\bigoplus}}{a^2(1-e^2)^{5/2}}\left[n_2^2\left(2-15\cos^2{i}\right)+10n_1n_2\cos i -n_1^2\right].
$$
The width computed in this way turns out to be very narrow. In Figure \ref{fig:width_high}, we provide an example for $A/m=1$ m$^2/$kg, with a close-up in the regions where the width is the largest. Notice that they correspond to resonances \#1 and \#3.

With respect to a possible ranking of the six resonances in terms of
the effect they might produce, the maximum eccentricity variation that
can be achieved at a resonance $j$ can be estimated by integrating the
first equation in (\ref{equations}). This is,
\begin{equation}\label{delta_e}
\Delta e = \left |\frac{3}{2}PC_R\frac{A}{m}\frac{\sqrt{1-e^2}}{na}\frac{\mathcal{T}_{j}}{\dot\psi_j}\right|.
\end{equation}
In Fig.~\ref{fig:delta_e} we show the corresponding values as a function of $i,a$ for $e=0.01$, $\Omega=0^{\circ}$, $\omega=0^{\circ}$, $\lambda_S=90.086^{\circ}$ and $A/m=1$ m$^2/$kg and $A/m=0.012$ m$^2/$kg, sampling the $(i,a)$ plane at a step of $\Delta i= 0.2^{\circ}$ and $\Delta a=20$ km. In the former case, the color bar representing the maximum eccentricity variation achievable is bounded to $0.3$, which is the value required to reenter from the highest LEO considered. In the latter case, the limit is set to $0.1$ just for the sake of clarity, that is, to be able to appreciate different behaviours, given the extreme narrowness of the resonances. 
 For $A/m=0.012$ m$^2/$kg, it should be noted that the variation that can be obtained is weak, and that, by definition of resonance, the time to get to the peak value is significantly long. Also, given that the range giving the required combination of inclination and semi-major axis is tapered, a possible exploitation of these corridors for an operational satellite would require an accurate manoeuvring capacity. A similar argument applies for resonance $\#6$ for $A/m=1$ m$^2/$kg.

 Note also that the same expression, Eq.~(\ref{delta_e}), can be used to quantify the maximum possible change in eccentricity in the neighborhood of a given resonance, eventually avoiding an unrealistic operational time. This application is faced in detail in \cite{Schettino_IAC2017}. We remark here that, for the strongest resonances (\#1, \#2 and \#3), the range of initial inclinations that can be targeted to achieve a considerable change in eccentricity for the high area-to-mass case is generally at most $\pm 1^{\circ}$ with respect to the resonant value $i^*$, for given initial semi-major axis and eccentricity.

\section{Numerical Evidence}

The analysis described in the previous Section was compared to the results of the numerical simulations performed under the ReDSHIFT H2020 project \citep{aR17, A2017a, A2017b}. The aim of the simulations was to map the whole LEO region in terms of initial orbital elements, considering two initial epochs and the same two values of area-to-mass ratio considered here, in order to look for natural deorbiting solutions that may facilitate the compliance with the international mitigation guidelines. A well-defined grid of initial conditions was propagated for 120 years by means of the semi-analytical orbital propagator FOP \citep{lAesa, aR09}, including in the dynamical model the geopotential up to degree and order 5, SRP, lunisolar perturbations and atmospheric drag below an altitude of 1500 km\footnote{The atmospheric model is based on a Jacchia-Roberts density model with an
exospheric temperature of 1000 K and a variable solar flux at 2800 MHz
(obtained by means of a Fourier analysis of data corresponding to the
interval 1961-1992).}. The grid in initial inclination, in particular, was set at steps of $2^{\circ}$ in the range $[2^{\circ}:120^{\circ}]$. 

\begin{figure}
\includegraphics[width=\columnwidth]{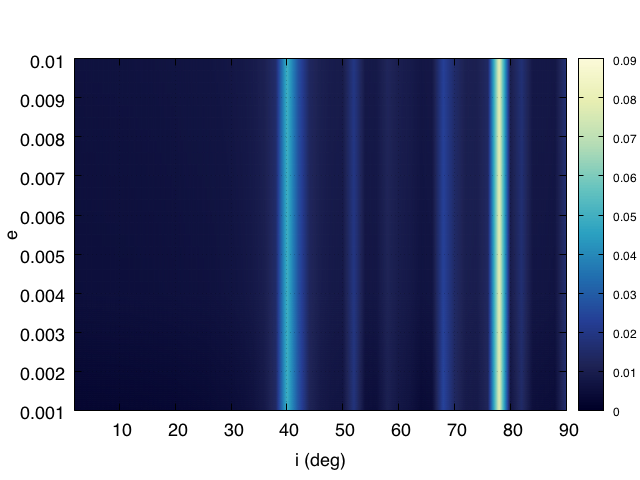}
\includegraphics[width=\columnwidth]{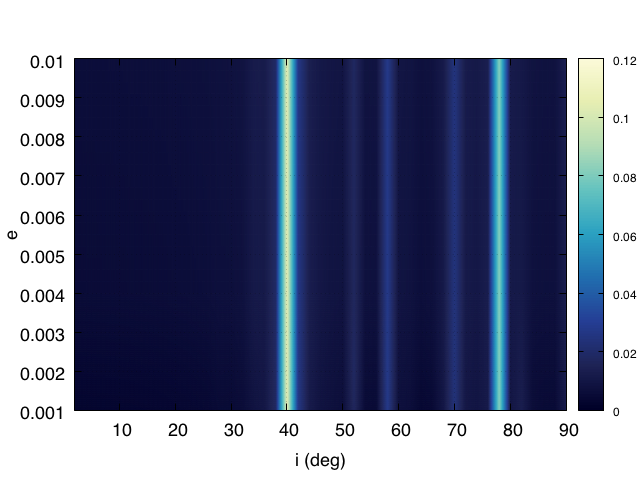}
\includegraphics[width=\columnwidth]{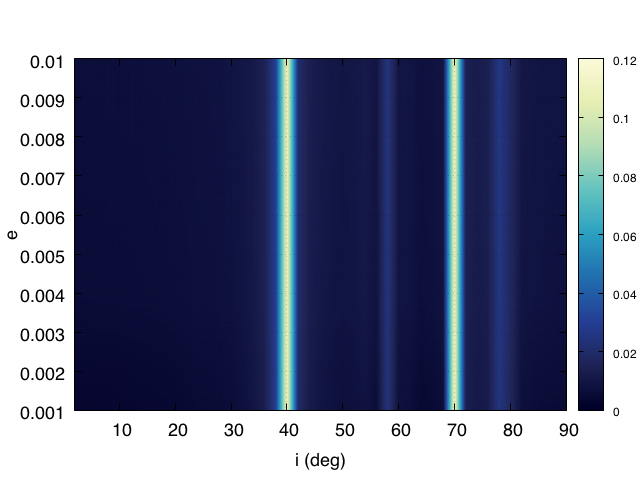}
    \caption{Maximum eccentricity (color bar) computed over 120 years with FOP, as a function of the initial inclination and eccentricity for $A/m=1$ m$^2/$kg, starting from $\Omega=0^{\circ}$, $\omega=0^{\circ}$ on June 21, 2020 at 06:43:12 (i.e., $\lambda_S\approx 90.086^{\circ}$). Top: $a=r_{\bigoplus} +1200$ km; middle:   $a=r_{\bigoplus} +1360$ km; bottom: $a=r_{\bigoplus} +1580$ km.}
   \label{fig:FOP}
\end{figure}

In Fig.~\ref{fig:FOP}, we show the maximum eccentricity achieved in 120 years, as a function of initial inclination and eccentricity, starting from three different values of semi-major axis ($a=r_{\bigoplus} +1200$ km, $a=r_{\bigoplus} +1360$ km, $a=r_{\bigoplus} +1580$ km), for quasi-circular orbits, with a same initial configuration, and assuming $A/m=1$ m$^2/$kg. It is possible to appreciate that the eccentricity can experience a non-negligible variation, only in correspondence of well-defined inclination bands (the brighter ones). These inclinations are located in the neighbourhood of the resonant values, described in the previous Section. In other words, within the limit of the grid adopted, only at resonant inclination values, the eccentricity can naturally increase. 

It is also interesting to note that in the three examples the role of the dominant resonance, that is the one providing the greatest variation, changes. This is what we have called  a {\it resonance switching}. In the top panel, this role is taken by resonance \#2; in the middle by resonance \#1; in the bottom by resonances \#1 and \#3. 

\begin{figure}
\includegraphics[width=\columnwidth]{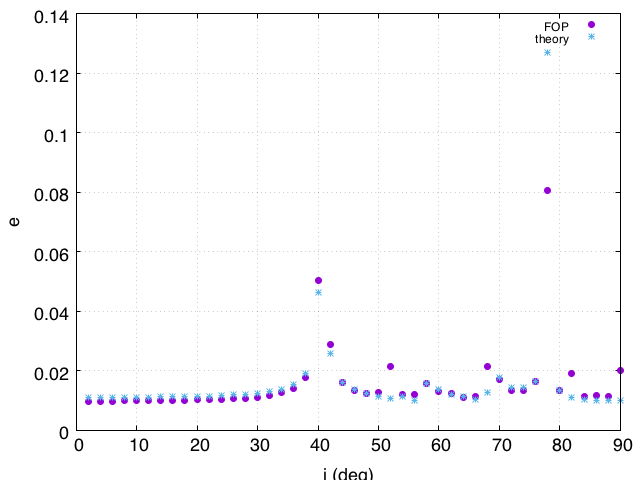}
\includegraphics[width=\columnwidth]{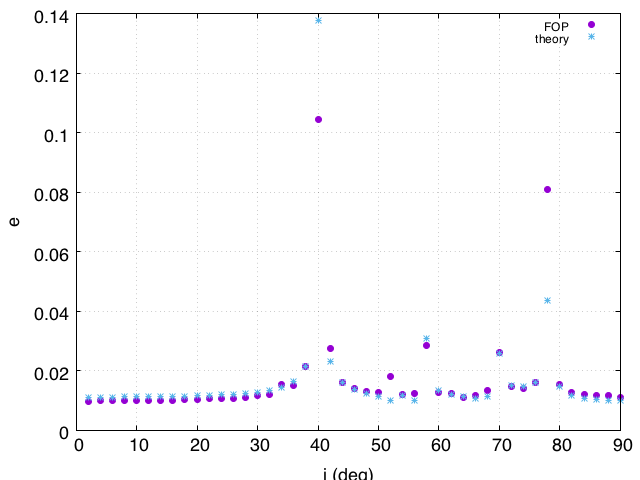}
\includegraphics[width=\columnwidth]{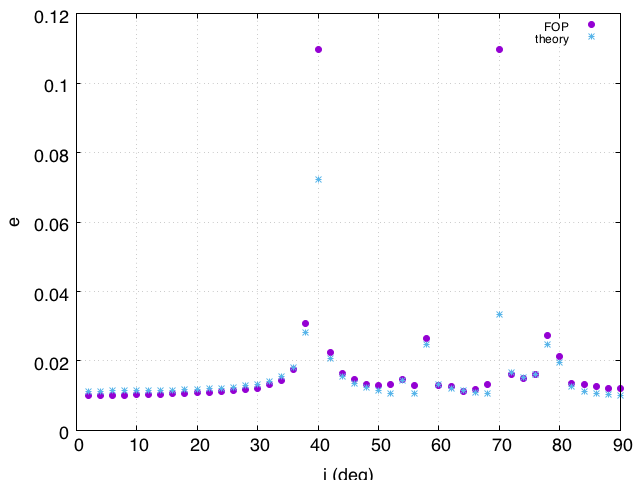}
    \caption{Maximum eccentricity computed over 120 years with FOP in purple, and following Eq.~(\ref{delta_e}) in cyan, as a function of the initial inclination, for $A/m=1$ m$^2/$kg, starting from $e=0.01$, $\Omega=0^{\circ}$, $\omega=0^{\circ}$ on June 21, 2020 at 06:43:12 (i.e., $\lambda_S\approx 90.086^{\circ}$). Top: $a=r_{\bigoplus} +1200$ km; middle:   $a=r_{\bigoplus} +1360$ km; bottom: $a=r_{\bigoplus} +1580$ km.}
   \label{fig:theoryFOP}
\end{figure}

In Fig.~\ref{fig:theoryFOP}, we compare the maximum eccentricity computed with the numerical simulation with the maximum variation obtained using estimate (\ref{delta_e}), for the same three initial conditions as in Fig.~\ref{fig:FOP}, considering $e=0.01$ as initial eccentricity. The analytical value depicted is the sum of the six variations forecast by Eq.~(\ref{delta_e}) for each resonance. This choice turned out to approximate better the actual behaviour of the maximum eccentricity, especially in the inclination regions between two resonant values. Given that in the simulations carried out with FOP other perturbations played a role and that the atmospheric drag was included, we consider that there exists a reasonable consistency between the numerical model and the simplified estimation. The resonance switching can also find an explanation, looking to the relative height corresponding to the cyan points, in particular for the first two examples.

\begin{figure*}
\includegraphics[width=\columnwidth]{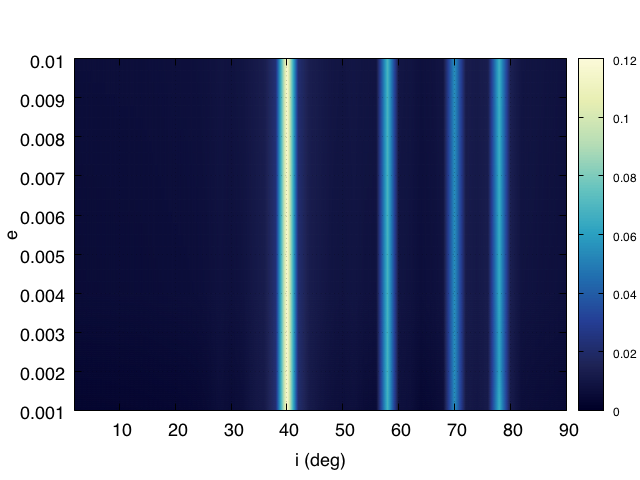}
\includegraphics[width=\columnwidth]{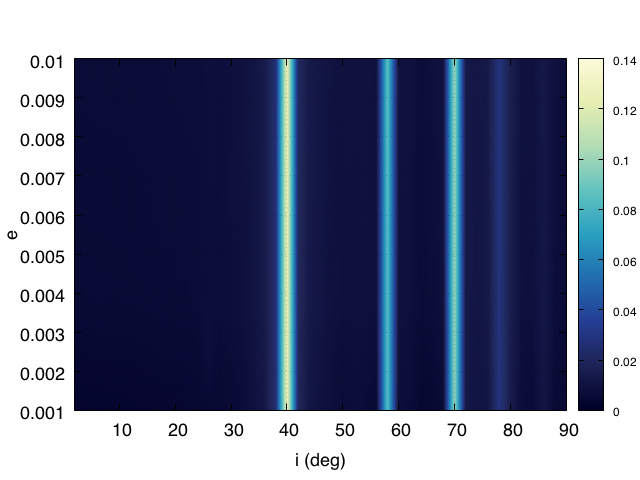}
    \caption{Maximum eccentricity (color bar) computed over 120 years with FOP, as a function of the initial inclination and eccentricity for $A/m=1$ m$^2/$kg, starting from $a=r_{\bigoplus} +1460$ km (left) and $a=r_{\bigoplus} +1520$ km (right). In both cases, $\Omega=0^{\circ}$, $\omega=180^{\circ}$ on June 21, 2020 at 06:43:12 (i.e., $\lambda_S\approx 90.086^{\circ}$). }
   \label{fig:14601520}
\end{figure*}

In general, for all the cases explored, it turned out that also resonance \#4 can yield a significant eccentricity variation, for the high area-to-mass value. See Fig.~\ref{fig:14601520} for two examples.

\begin{figure*}
\includegraphics[width=\columnwidth]{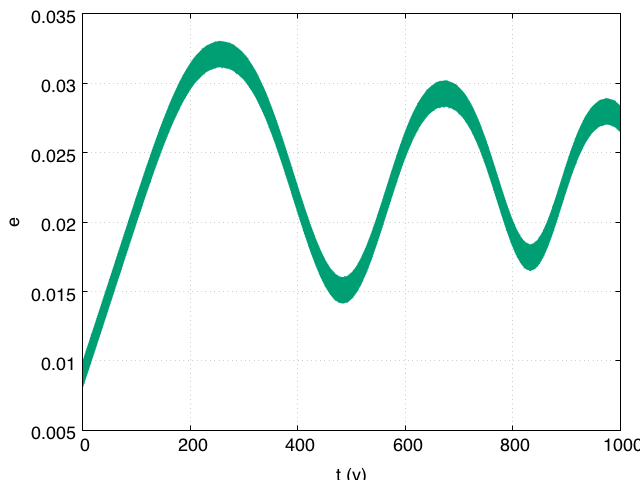}
\includegraphics[width=\columnwidth]{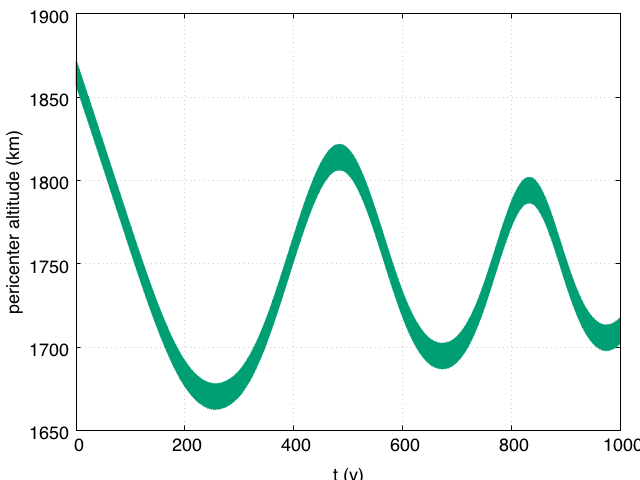}
    \caption{On the left,  eccentricity evolution computed with FOP for $A/m=0.012$ m$^2/$kg, starting from $a=r_{\bigoplus} +1940$ km, $e=0.01$, $i=80^{\circ}$, $\Omega=90^{\circ}$, $\omega=90^{\circ}$ on June 21, 2020 at 06:43:12 (i.e., $\lambda_S\approx 90.086^{\circ}$). On the right, the consequent variation in pericenter altitude. }
   \label{fig:80}
\end{figure*}

\begin{figure*}
\includegraphics[width=\columnwidth]{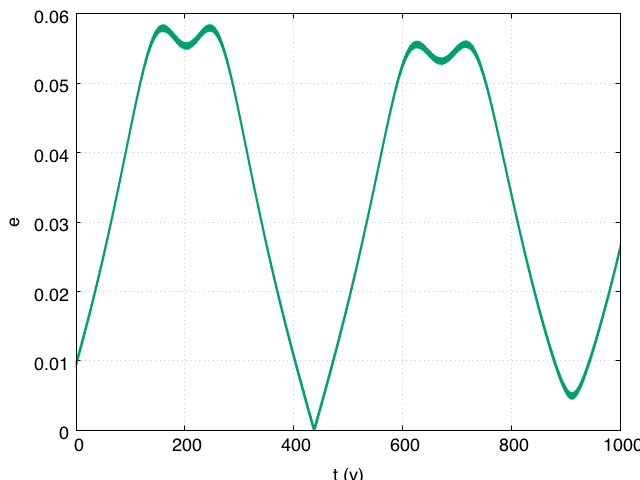}
\includegraphics[width=\columnwidth]{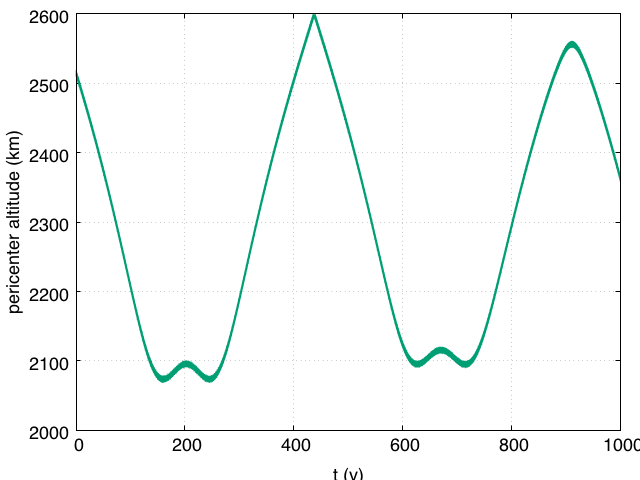}
    \caption{On the left,  eccentricity evolution computed with FOP  for $A/m=0.012$ m$^2/$kg, starting from $a=r_{\bigoplus} +2600$ km, $e=0.01$, $i=36^{\circ}$, $\Omega=90^{\circ}$, $\omega=90^{\circ}$ on June 21, 2020 at 06:43:12 (i.e., $\lambda_S\approx 90.086^{\circ}$). On the right, the consequent variation in pericenter altitude. }
   \label{fig:40}
\end{figure*}

\begin{figure}
\includegraphics[width=\columnwidth]{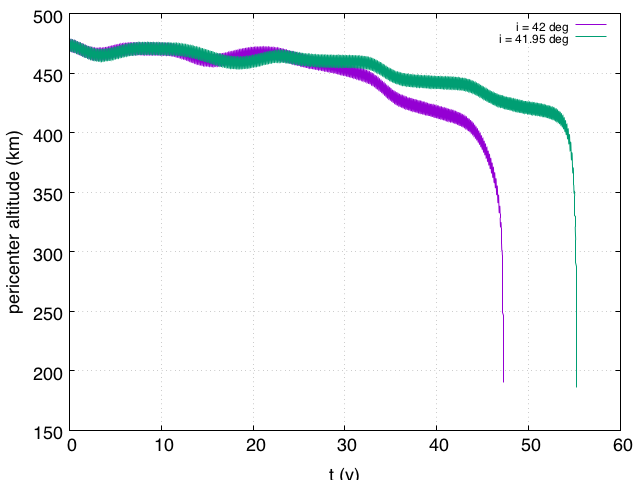}
    \caption{Pericenter altitude evolution computed with FOP for $A/m=0.012$ m$^2/$kg, starting from $a=r_{\bigoplus} +840$ km, $e=0.05$, $i=42^{\circ}$ (purple) and $i=41.95^{\circ}$ (green), $\Omega=270^{\circ}$, $\omega=270^{\circ}$ on June 21, 2020 at 06:43:12 (i.e., $\lambda_S\approx 90.086^{\circ}$). }
   \label{fig:lowAM42}
\end{figure}

For $A/m=0.012$ m$^2/$kg, in Figs.~\ref{fig:80}--\ref{fig:40} we provide two examples of the eccentricity evolution and the consequent variation in the pericenter altitude by propagation with FOP, for altitudes where the atmospheric drag is not effective. They correspond to quasi-circular orbits, which are located at the resonance $\#2$ and $\#1$, respectively, at the initial epoch. We note the variation of hundreds of km obtained, but also the long time required. In Fig.~\ref{fig:lowAM42}, we show an example when the SRP resonance aids the action of the atmospheric drag. Starting from $i=41.95^{\circ}$, the reentry is achieved in about 55 years; starting from $i=42^{\circ}$, the lifetime is reduced by about 8 years. This shows that if the satellite is located in the neighbourhood of a SRP resonance\footnote{This is, the change in velocity to target the specific inclination can be afforded.}, it could be worth to increase a little the eccentricity to clearing the region more rapidly. In this way, we can take advantage of both putting the spacecraft in a denser layer of the atmosphere and the push in eccentricity given by SRP, which can lower the pericenter altitude in the long-term. An accurate modelling of the interaction between SRP and atmospheric drag will be faced in the future.

\section{Conclusions}

In this work, we have reviewed the main orbital resonances associated with solar radiation pressure which affects the evolution in eccentricity, in order to see if they might be exploited to design advantageous deorbiting solutions for Low Earth Orbits. Starting from a well-known theory, we have showed that at least four out of six resonances can be considered to increase the eccentricity of a small spacecraft equipped with a solar sail. The variation that can be achieved along a given resonance, that is, at well-defined values of semi-major axis, inclination  and eccentricity, is such that a reentry to the Earth can be obtained, even from high altitudes. For typical values of operational satellites, SRP resonances can be considered, instead, in combination with the atmospheric drag, provided  an accurate manoeuvring capability. The simplified model including only solar radiation pressure and Earth oblateness has been validated against a more realistic dynamical model. The corresponding results turn out to be consistent. The novelty of the work regards the possible applications, but it is also theoretical, because we have shown the importance of resonances which were neglected in the past.

Future work will look into the role of the initial phasing provided by initial epoch, longitude of ascending node and argument of pericenter, which could ensure either an increase or an eccentricity decrease. A detailed analysis on the characteristic frequencies for the eccentricity evolution in the resonant regions due to SRP is ongoing. That study will include also a comparison between the theoretical derivation explained here and the numerical results, without the contribution of the drag. Part of the results can be found in \cite{Schettino_IAC2017}. Also, the inclination evolution was neglected in the analysis presented, and it will be evaluated in detail. From the point of view of a space operator, the model is still simplified in the sense that the orientation of a sail cannot be assumed as always directed normal to the solar radiation. This is another issue, which will deserve further research.

\section*{Acknowledgements}

This work is funded through the European Commission Horizon 2020, Framework
Programme for Research and Innovation (2014-2020), under the ReDSHIFT project
(grant agreement n$^{\circ}$ 687500).

We are grateful to Bruno Carli, Camilla Colombo and Kleomenis Tsiganis for the useful discussions we had. We thank the anonymous reviewer for the comments that improved considerable the paper. E.M.A. is also grateful to Florent Deleflie for the period spent in Lille, and to Jerome Daquin for the suggestions given. 








%
%


\bsp	
\label{lastpage}
\end{document}